\documentclass[aps,prb,twocolumn,showpacs,amsmath,amssymb]{revtex4}
\usepackage{graphicx}
\usepackage{dcolumn}
\usepackage{bm}
\usepackage{color}
\usepackage{graphicx}
\usepackage{epsfig}
\include{graphic}

\begin{document}

\title{High sensitivity microwave detection using a magnetic tunnel junction in the absence of an external applied magnetic field}

\author{Y. S. Gui$^{1}$, Y. Xiao$^2$, L. H. Bai$^1$, S. Hemour$^3$, Y. P. Zhao$^3$, D. Houssameddine$^4$, K. Wu$^3$, H. Guo$^2$, and C.-M. Hu$^1$}

\affiliation{$^{1}$Department of Physics and Astronomy, University
of Manitoba, Winnipeg R3T 2N2, Canada }

\affiliation{$^{2}$ Department of Physics, Center for the Physics of Materials, McGill University, Montreal, Quebec H3A 2T8, Canada}

\affiliation{$^{3}$Ecole Polytechnique de Montral, Montral H3T 1J4, Canada }

\affiliation{$^{4}$Everspin Technologies, 1347 N. Alma School Road, Chandler, Arizona 85224, USA}


\begin{abstract}

In the absence of any external applied magnetic field, we have found that a magnetic tunnel junction (MTJ) can produce a significant output direct voltage under microwave radiation at frequencies, which are far from the ferromagnetic resonance condition, and this voltage signal can be increase by at least an order of magnitude by applying a direct current bias. The enhancement of the microwave detection can be explained by the nonlinear resistance/conductance of the MTJs. Our estimation suggests that optimized MTJs should achieve sensitivities for non-resonant broadband microwave detection of about 5,000 mV/mW.

\end{abstract}


\maketitle


Magnetic tunnel junctions (MTJs) are one of the most important structures in the field of spintronic, based on their static properties; MTJs are currently widely used for read-heads of hard disk drives and non-volatile memory (magnetoresistive random-access memory, MRAM).\cite{Chappert2007} Recent dynamics studies of MTJs have revealed new principles for devices such as nano-oscillators and spin diodes and thus opened many possibilities to incorporate MTJ devices in microwave applications. \cite{Tulapurkar_Spindiode, Houssameddine2009, Sun2008, Slavin2009, Fan2009} The use of spin diode effect caused by spin transfer torque, \cite{STT1, STT2} aids the development of a new generation of microwave detector based on MTJ devices.\cite{Tulapurkar_Spindiode, Wang2009, Ishibashi2010,Prokopenko2013, Miwa2014, Fang2014} As a consequence, the sensitivity of spintronic microwave detectors, which is characterized by the ratio between the produced dc voltage and the incident microwave power,\cite{Hemour2014} has been improved by more than four orders of magnitude in a decade,  from the earliest report\cite{Tulapurkar_Spindiode} of 1.4 mV/mW  to the latest report \cite{Fang2014} of 74,500 mV/mW, that can compete with the commercial semiconductor Schottky diode detectors. Since the nano-structured MTJ mcirowave sensor has a small size compared with the conventional diode detector it should cause less perturbation of the microwave field under test. Besides the smaller size and overcoming the limitation caused by the thermal voltage \cite{Hemour2014}, it has been found that in spintronic microwave detector the nonlinear effect enhances the signal more than the noise as the size of the magnets decreases.\cite{Miwa2014} This indicates that a spintronics device may overcome the limitations of semiconductor devices and eventually reach thermo-dynamic limits.\cite{Miwa2014}

Note that the conventional semiconductor microwave detector as well as the novel integrated photonic microwave sensor, which has been recently developed in a broad frequency range with high sensitivity and achieved a minimum detectable power density of 8.4 mW/m$^2$,\cite{Zhang2014, Savchenkov2014} do not require a magnetic bias. Thus, the non-resonant microwave rectification in both ferromagnetic single layer \cite{Zhu2011} and multilayer structures \cite{Zhang2012, Fu2012, Fu2014} has also been developed; however, the sensitivity is on the order of 1-10 mV/mW, which is only comparable to early reports of resonant sensitivity.\cite{Tulapurkar_Spindiode} In this paper, we demonstrate that the non-resonant microwave response can be significantly enhanced by an appropriate DC current bias in addition to the microwave radiation. The sensitivity of an MTJ at zero magnetic field can be increased up to 350 mV/mW by applying a small DC current bias, about one order of magnitude improvement compared with that at zero current bias. It should be noted that the  enhancement occurs at microwave frequencies far from the ferromagnetic resonance condition, where the spin-torque diode effect can be neglected because the cone angle of magnetization precession tends to be zero. To explore the understanding of non-resonant rectification effects, we propose a phenomenological model based on the nonlinear resistance/conductance of the MTJs, which can explain the features of the experimental observation.

\begin{figure} [t]
\begin{center}
\epsfig{file=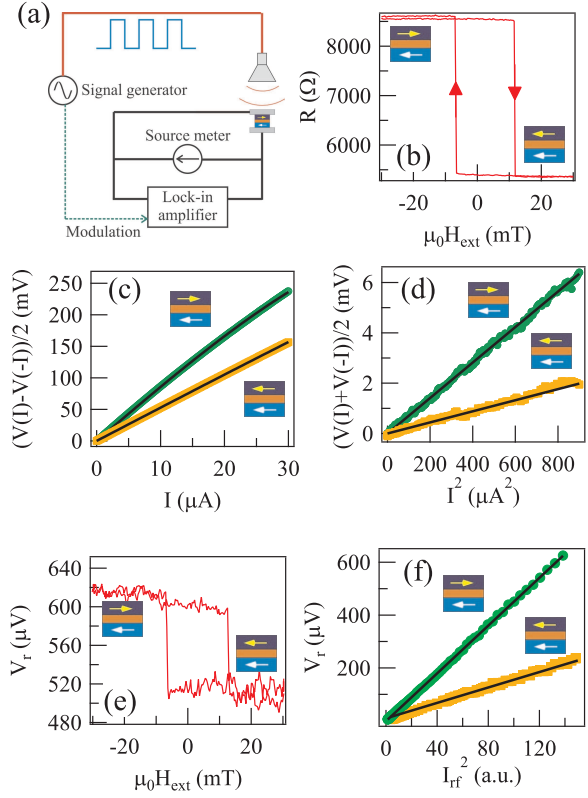,width=8cm} 
\caption{(color online) (a) Diagram of experimental setup for microwave measurement. (b) The resistance of an MTJ as a function of the magnetic field and its sweeping direction with a current bias of 10 $\mu$A and without microwave radiation. The odd component (c) and even component (d) of $I-V$ curves as functions of $I$, where solid lines are fittings based on Eq.(\ref{Eq:dcIV}). (e) Magneto-dependent $V_r$ measured at 6.8 GHz. The output power of the signal generator is 10 dBm. (f) $V_r$ (symbols) as function of $I_{r\!f}^2$, with solid lines indicating a linear fitting. }\label{Fig1}
\end{center}
\end{figure}

The primary experimental setup is shown in Fig. \ref{Fig1}(a); here a microwave current, $I_{r\!f}$, can be produced in an MTJ device under microwave radiation using a signal generator (Anritsu, 3692C) and a horn antenna, while a DC bias current, $I$, is applied by a source-meter (Keithley, 2400). This setup effectively isolates the DC measurement system from the high frequency circuit, and will also show that the MTJ device is capable of receiving external microwave signals. Pulse modulated microwaves and a lock-in amplifier (Stanford Research Systems, SRS830) are used to separate the rectified voltage due to the microwave radiation from the voltage solely caused by the DC current. 

The MTJ film with an RA product of about 9.5 $\Omega\mu{}m^2$ is grown on a Si substrate covered with 200 nm SiO$_2$ and the 1.2 nm MgO tunnel barrier is formed by sputtering of Mg followed by an oxidation step. The tunneling magnetoresistance (TMR) ratio is about 62\%. The hard magnetic layer of the MTJ is a complex multilayer system composed of a CoFe(2.3nm)/Ru(0.8nm)/CoFeB(2.2nm)/CoFe(0.5nm) synthetic antiferromagnetic structure pinned by exchange bias using PtMn (20nm). The soft magnetic layer, buffer, and capping layers are CoFeB(2.5 nm), TaN and Ta, respectively. The MTJ stack is patterned into an elliptical shape with short axes from 63 nm to 120 nm and the aspect ratio from 1.8 to 3.5. The configuration of magnetization can be switched by a small in-plane magnetic field ($H_{ext}$) of about $\pm$25 mT; single domain magnetization reversal for an elliptical sample with long and short axes of 145 and 63 nm, respectively is seen in Fig.\ref{Fig1}(b).

Starting from the DC transport measurements for MTJ devices without microwave radiation, the voltage across the junction includes contributions from higher order terms besides the ohmic term ($R_0I$), expressed as

\begin{equation}
V(I)=I(R_0+\xi{}I+\eta{}I^2+\cdots),
\label{Eq:dcIV}
\end{equation}
    
\noindent where $R_0$ is the ohmic resistance. At low current bias, only the linear ohmic ($R_0$),  quadratic ($\xi$) and cubic ($\eta$) terms are taken into account to simplify the mathematical expression, which can well explain the measurement results as discussed in detail in the following text. It should be noted that higher order $I$ terms may contribute significantly at large currents.

We first perform DC measurements for MTJ devices and extract the important parameters of $R_0$, $\xi$, and $\eta$. The I-V measurements have been performed at either a parallel state (PS) or anti-parallel state (APS). Changing the polarization of the current $I$, the odd and even parts of the $I-V$ curves can be determined as $(V_+-V_-)/2=R_0I+\eta{}I^3$ and $(V_++V_-)/2=\xi{}I^2$. As shown in Fig.\ref{Fig1}(c), the contribution of $\eta$ is significant and results in a deviation of $(V_+-V_-)/2$ from the linear relation. The parameters $R_0$, $\xi$ and $\eta$ are deduced to be $R_0^{AP}=8650$ $\Omega$, $\xi^{AP}=7.1\times10^6$ V/A$^2$, and $\eta^{AP}=-8.7\times10^{11}$ V/A$^3$ for the APS, and $R_0^P=5270$ $\Omega$, $\xi^P=2.2\times10^6$ V/A$^2$, and $\eta^{P}=-5.5\times10^{10}$ V/A$^3$ for the PS. Note that both $\xi$ and $\eta$ are strongly dependent on the magnetization configuration in the MTJ, as is $R_0$, which is so far the basis for the application of MTJ devices.  

When a microwave current $I_{r\!f}\sin(\omega{}t)$ instead of a DC current $I$ is flowing in the device, a microwave photovoltage  ($V_r=\xi{}I_{r\!f}^2/2$) is generated across the junction as shown Fig. \ref{Fig1}(e), which is linearly dependent on $I_{r\!f}^2$ (Fig. \ref{Fig1}(f)) and hence dependent on the incident microwave power. Our preliminary results indicate a frequency range from 1 MHz to 40 GHz for the MTJ based microwave detector.\cite{Hemour2014} Using the lock-in technique, the minimum detectable voltage signal is about 100 nV limited by the noise level and the maximum measured voltage signal can be as high as 10 mV at 6.8 GHz. These values allow us to further estimate the minimum and maximum microwave power density \cite{Zhang2014, Pozar2011} of about 10 mW/m$^{2}$ and 1$\times$10$^{6}$ mW/m$^{2}$, respectively. Multiple mechanisms such as spintronic\cite{Tulapurkar_Spindiode}, electronic\cite{Brinkman} and thermoelectric \cite{Zhang2012} effects may contribute to the nonlinear $\xi$ term, depending on the particular sample structure and material properties. Theoretically, comprehensive insight into the nonlinear behaviour in an MTJ and the resultant microwave response are still under development. Experimentally, at GHz frequencies the broadband microwave measurement enables to distinguish resonant and non-resonant effects occurring at $H_{ext}=0$,  and the combination of external laser heating and internal microwave heating is an effective method to distinguish nonlinear transport caused by thermoelectric and electronic effects\cite{Zhang2012}. 

Microwave detectors are often operated with a DC bias; therefore, for generality, the DC current bias should be included. By including the superposition of the DC current bias $I$ and the microwave current $I_{r\!f}\sin(\omega{t})$, one can calculate the voltage signal biased by both the DC and microwave current as

\begin{eqnarray}
\nonumber 
V(I,I_{r\!f},t)&=&(\frac{1}{2}\xi{}+\frac{3}{2}\eta{}I)I_{r\!f}^2+(R_0I+\xi{}I^2+\eta{}I^3)\\
\nonumber      &&+(R_0+2\xi{}I+3\eta{}I^2+\frac{3}{4}\eta{}I_{r\!f}^2)I_{r\!f}\sin(\omega{t})\\
\nonumber      &&-\frac{1}{2}(\xi+3\eta{}I)I_{r\!f}^2\cos(2\omega{t})-\frac{1}         {4}\eta{}I_{r\!f}^3\sin(3\omega{t}).\\
       \label{Eq:acdcIV}      
\end{eqnarray}

It is clearly seen that the additional DC voltage produced by the microwave field is $\xi{}I_{r\!f}^2/2+3\eta{}II_{r\!f}^2/2$, including not only the photo-voltage $PV\equiv\xi{}I_{r\!f}^2/2$ but also the contribution of photo-resistance $PR\equiv3\eta{}I_{r\!f}^2/2$, which was previously used to calibrate $I_{r\!f}^2$ because it is linearly proportional to microwave power ($\propto{I_{rf}^2}$).\cite{Sankey2008}  It is believed that the microwave magnetic field may slightly tilt the magnetization from its equilibrium direction and hence change the resistance of an MTJ. Since the resistance/conductance of an MTJ relies on spin polarized quantum tunneling, a detailed analysis of the photo-resistance effect should have to use the quantum coherent transport model which will be discussed elsewhere.\cite{Xiaopreprint}

The voltage responsivity of a microwave detector is defined as the ratio of the produced DC voltage over the average microwave power absorbed by the device, and can be determined from the I-V relation [Eq. (\ref{Eq:dcIV})] as \cite{Rv1, Rv2}

\begin{equation}
R_v\equiv\frac{dV^2/dI^2}{2dV/dI}=\frac{\xi+3\eta{}I}{R_0+2\xi{I}+3\eta{I^2}}.
\label{Eq:sensitivity}
\end{equation}

At zero current bias, the intrinsic voltage responsivity is solely determined by the ratio of $\xi$ over $R_0$. Applying an appropriate DC bias, the voltage responsivity of an MTJ microwave detector can be significantly enhanced because of the coupling between the photo-resistance $PR$ and the applied DC current, expressed as the $3\eta{}I$ term in the numerator of Eq. (\ref{Eq:sensitivity}). Regarding the fact that a portion of the microwave power absorbed by the device is inevitably dissipated in the parasitic elements, the measured sensitivity, $\varepsilon$, of MTJ microwave detectors at GHz range may differ from $R_v$ in Eq. (\ref{Eq:sensitivity}).

\begin{figure} [t]
\begin{center}
\epsfig{file=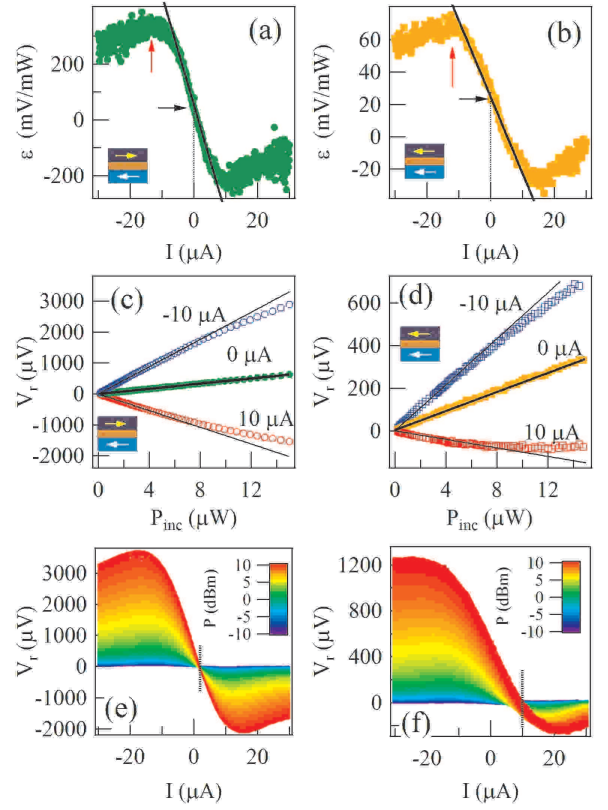,width=8 cm}
\caption{(color online) Bias DC current dependence of the sensitivity of the MTJ microwave detector at $P_{inc}$=0.15 $\mu{}W$ for APS (a) and PS (b), respectively. Dotted lines indicate the sensitivity at $I=0$ and red arrows indicate the enhancement of sensitivity. Solid lines indicate the linear dependence at low current bias. Rectified voltage as a function of incident microwave power $P_{inc}$ in the MTJ at several DC current biases for APS (c) and PS (d), respectively.  Solid lines follow linear fittings at lower $P_{inc}$. (e) and (f) Microwave rectification as a function of DC bias for different output powers from -10 dBm to 10 dBm for APS and PS, respectively. The dotted line indicates the critical DC bias for $V_r=0$. The microwave frequency is 6.8 GHz.}\label{Fig2}
\end{center}
\end{figure}

Now we perform lock-in measurements to quantitatively determine the sensitivity of an MTJ microwave detector under DC current bias. As shown in Fig. \ref{Fig2}(a), the sensitivity is about $\varepsilon=40$ mV/mW at $I$=0 for APS after the power calibration, indicated by the dotted line and black arrow. A significant enhancement of sensitivity appears at negative DC currents, with $\varepsilon=350$ mV/mW at $I=-14$ $\mu$A (indicated by the red arrow), dominated by the PR effect. The sensitivity at $I=14$ $\mu$A is also larger with a value of $\varepsilon=-260$ mV/mW. At low DC current bias(8 $\mu$A$>$I$>$-8 $\mu$A), the sensitivity is linearly dependent on $I$ as indicated by the solid line, which is in agreement with the expected linear dependence on $I$ according to Eq.(\ref{Eq:sensitivity}). The negative slope in Fig. \ref{Fig2}(a) can be attributed to the fact that $\eta<0$. Similar dependence of the sensitivity on DC current bias are also observed for PS as shown in Fig. \ref{Fig2}(b), Since $\eta^{P}/\eta^{AP}\ll{}1$ the enhancement of the sensitivity for PS is not as significant as that for APS and the resultant maximum sensitivity for PS is only about 70 mV/mW  (one fifth of that for APS) as indicated by the red arrow. 

The sensitivity amplitude tends to slowly decrease at larger DC current biases (either positive or negative) for both APS and PS. We note that the observation is almost unchanged as the microwave power is increased from $P=-10$ dBm (corresponding to an incident microwave power of $P_{inc}=0.15$ $\mu$W) to $P=10$ dBm (corresponding to $P_{inc}=15$ $\mu$W). The decrease in sensitivity with a larger DC current is also seen in an MTJ microwave detector at ferromagnetic resonance\cite{Ishibashi2011}, which may attribute to the decrease of TMR at a large current bias.

At zero current bias, the rectified voltage $V_r$ measured by the lock-in amplifier shows a linear dependence on the incident microwave power for APS in Fig. \ref{Fig2}(c) and PS in Fig. \ref{Fig2}(d), respectively, indicating a constant sensitivity for both the states up to 15 $\mu$W. At a DC current bias of $\pm10$ $\mu$A, $V_r$ originates from the linear dependence and the sensitivity tends to decrease when $P_{inc}>8$  $\mu$W.

We note that the measured sensitivity of 40 mV/mW at zero current bias is much smaller than the intrinsic responsivity \cite{Rv1, Rv2} $R_v=\xi/R_0$ of about $800$ mV/mW deduced from DC transport experiments. While a more realistic circuit model is still under development, the deduction can be explained by a classical model with the parasitic series resistance $R_S$ and barrier capacitance $C_B$, which results in a factor of $1/(1+R_S/R_0)[1+(f/f_c)]^2$, where $f_c=\sqrt{1+R_s/R_0}/2\pi{}C_B\sqrt{R_SR_0}$ is the characterization frequency and $f$ is the imposed microwave frequency.\cite{Rv1} The measured sensitivity at 20 MHz, i.e. $f\ll{}f_c$, of about 500 mV/mW, is in agreement with this estimation. This effect implies that the sensitivity at zero current bias could exceed 500 mV/mW in the GHz range by decreasing the parasitic elements. Supposing a similar enhancement due to DC current bias, a sensitivity of about 5,000 mV/mW is achievable for MTJ microwave detectors in the absence of magnetic fields.  

Moreover, Eq. (\ref{Eq:sensitivity}) predicts that the sensitivity is zero at a critical DC current $I^c=-\xi/3\eta$, where the produced $V_r$ is zero no matter how strong the microwave current is. Indeed, this effect has been observed as shown in Fig. \ref{Fig2}(e) for APS, where a crossing point at $I^c=2$ $\mu$A (indicated by the dotted line) is clearly seen despite the radiated microwave power varying over two orders of magnitude. A rough estimation based on $\xi^{AP}=7.1\times10^6$ V/A$^2$ and $\eta^{AP}=-8.7\times10^{11}$ V/A$^3$ from DC transport data results in a critical DC current of $I^c=2.7$ $\mu$A. The consistency of $I^c$ between our experiments and the phenomenological model is good keeping in mind the fact that no adjusted parameter was used in the estimation. As expected the critical point appears at high DC currents for PS as shown in Fig. \ref{Fig2}(f).

\begin{figure} [t]
\begin{center}
\epsfig{file=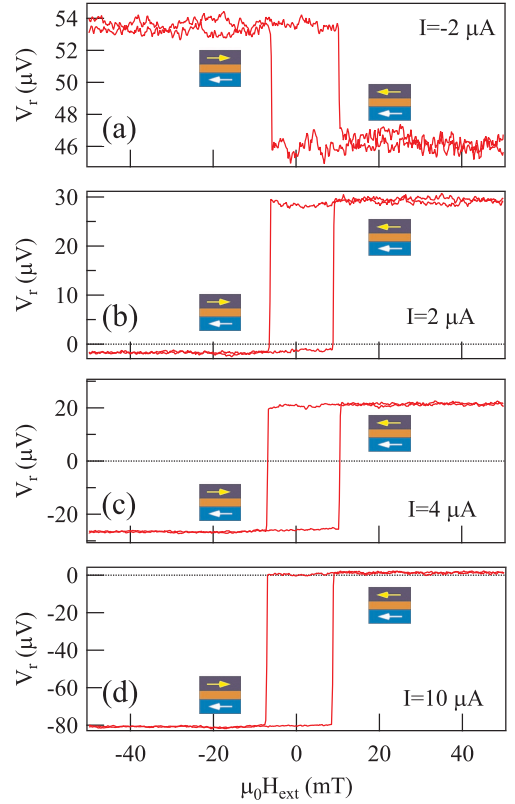,width=7 cm} 
\caption{(color online)  Magneto-dependent $V_r$ measured at 7.2 GHz at several biased DC currents. The output power of the signal generator is 3 dBm. Dottted lines indicate $V_r$=0.}\label{Fig3}
\end{center}
\end{figure}

Similar to the resistance of the MTJ devices, $V_r$ is also dependent on the magnetic configuration and hence could be used to read the information stored in the MTJ cell. To characterize the magneto-dependence of $V_r$, the dynamic tunneling magnetoresistance (DTMR) is calculated, which is defined as $\mathrm{DTMR}=|V_r^{APS}-V_r^{PS}|/\mathrm{min}(|V_r^{APS}|,|V_r^{PS}|)$ similar to the TMR effect $\mathrm{TMR}=|R_0^{APS}-R_0^{PS}|/\mathrm{min}(R_0^{APS},R_0^{PS})$, where $\mathrm{min}$ indicates the minimum value inside the brackets.  In the absence of the current bias, the $V_r$ loop observed as shown in Fig. \ref{Fig1}(e) has a smaller DTMR$\sim20\%$ as compared to TMR$\sim60\%$. However,  the DTMR can be significantly increased to a level extending to 10,000\%  by applying an external DC current (at $I$=10 $\mu$A) as shown in Fig. \ref{Fig3} since $V_r$ can be completely suppressed by an appropriate DC current bias for either APS or PS. 

Not only the magnitude of the $V_r$ is magnetically controlled, but also the polarity of $V_r$, which is beyond the TMR effect. At a negative current bias, $V_r$ for both APS and PS is always positive and a typical $V_r$-loop is plotted in Fig. \ref{Fig3}(a) at $I=$-2 $\mu{}A$. Increasing the current bias, $V_r$ decreases for both APS and PS, but more rapidly for APS. It is found $V_r^{APS}\sim0$ at $I=$2 $\mu{}A$ and $V_r^{PS}\sim0$ at $I=$10 $\mu{}A$ as shown in Fig. \ref{Fig3}(b) and (d), respectively. These results are consistent with the observations from Fig. \ref{Fig2}(e) and (f), thus, confirming the critical current $I^c$ is insensitive to both the microwave frequency and its power. Between 2 $\mu{}$A and 10 $\mu{}$A, the polarization of $V_r^{APS}$ and $V_r^{PS}$ is opposite as shown clearly in Fig. \ref{Fig3}(c) with $I=4$ $\mu{}A$. When $I>$ 10 $\mu$A, both $V_r^{APS}$ and $V_r^{PS}$ become negative.

In summary, we have investigated the sensitivity of MTJ microwave detectors in the absence of external applied magnetic field. It is found that a DC current of about -10 $\mu$A can enhance the sensitivity up to one order of magnitude and sensitivities of about 350 mV/mW are obtained. A phenomenological model including nonlinear resistance of the MTJ device can well explain this DC enhancement of AC transport in an MTJ device, thereby achieving useful understanding of the MTJ microwave detector.  It is believed that the sensitivity of MTJ microwave detectors may further be enhanced up to 5,000 mV/mW in the absence of any external applied magnetic field by optimizing the MTJ structure by including, e.g. (100) textured stacks with bcc ferromagnetic electrodes and MgO barrier and materials with large polarization and low damping, which lead to large spin polarization/TMR\cite{Zhang2004}. As specific examples of potential applications the optimized spintronic sensor could be used for ambient microwave energy harvesting, where conventional Schottky diodes fails to provide satisfactory microwave to DC conversion efficiency mainly because of its high zero-bias junction resistance, \cite{Hemour2014, Hemour}  as well as the non-destructive detection\cite{Fu2012} and microwave radar imaging\cite{Fu2014} since the small size of the MTJ microwave sensor would cause less perturbation of microwave under test.

We gratefully acknowledge financial support from Defence Research and Development Canada, under the contract W7701-135549/001/QCL, and NSERC grants. We would like to thank H. Abou-Rachid, Z. H. Zhang, P. Hyde and L. Fu for discussions.


\end{document}